\documentclass{aastex631}
\usepackage{amsmath}
\usepackage{hhline}
\usepackage{multirow} 
\usepackage{ragged2e} 

\shorttitle{Climate Bistability near the IEHZ}
\shortauthors{Fan et al.}

\graphicspath{{./}{figures/}}

\begin{document}

\title{Climate Bistability at the Inner Edge of the Habitable Zone due to Runaway Greenhouse and Cloud Feedbacks}

\correspondingauthor{Bowen Fan}
\email{bowen27@uchicago.edu}

\author[0000-0002-7652-9758]{Bowen Fan}
\affiliation{Department of the Geophysical Sciences, University of Chicago, Chicago, IL, 60637, USA}

\author[0000-0001-7180-6827]{Da Yang}
\affiliation{Department of the Geophysical Sciences, University of Chicago, Chicago, IL, 60637, USA}

\author[0000-0001-8335-6560]{Dorian S. Abbot}
\affiliation{Department of the Geophysical Sciences, University of Chicago, Chicago, IL, 60637, USA}

\begin{abstract}
Understanding the climate dynamics at the inner edge of the habitable zone (HZ) is crucial for predicting the habitability of rocky exoplanets. Previous studies using Global Climate Models (GCMs) have indicated that planets receiving high stellar flux can exhibit climate bifurcations, leading to bistability between a cold (temperate) and a hot (runaway) climate. However, the mechanism causing this bistability has not been fully explained, in part due to the difficulty associated with inferring mechanisms from small numbers of expensive numerical simulations in GCMs. In this study, we employ a two-column (dayside and nightside), two-layer climate model to investigate the physical mechanisms driving this bistability. Through mechanism-denial experiments, we demonstrate that the runaway greenhouse effect, coupled with a cloud feedback on either the dayside or nightside, leads to climate bistability. We also map out the parameters that control the location of the bifurcations and size of the bistability. This work identifies which mechanisms and GCM parameters control the stellar flux at which rocky planets are likely to retain a hot, thick atmosphere if they experience a hot start. This is critical for the prioritization of targets and interpretation of observations by the James Webb Space Telescope (JWST). Furthermore, our modeling framework can be extended to planets with different condensable species and cloud types.
\end{abstract}


\section{Introduction} \label{sec:intro}

The inner edge of the habitable zone is a crucial threshold in exoplanetary science, delineating the closest orbit around a star where a rocky planet can sustain liquid water on its surface \citep{kasting1993habitable, kopparapu2013habitable}, before a planet experiences a runaway \citep{komabayasi1967discrete, ingersoll1969runaway} or moist greenhouse \citep{kasting1993habitable}. Understanding the climate dynamics at this boundary is essential for assessing the habitability of exoplanets \citep{seager2013exoplanet}. The precise location of the inner edge of the HZ is influenced by various atmospheric processes, in particular the greenhouse effect and cloud feedbacks.

Early 3D GCM modeling \citep{yang2013stabilizing,yang2014low,yang2014strong} highlighted the significant role of dayside cloud feedbacks on tidally-locked and slowly-rotating planets. Their studies showed that increased cloud cover on the dayside can reflect more stellar radiation, thereby cooling the planet and extending the habitable zone closer to the star than indicated by previous 1D single-column models. This work has been followed up by other groups \citep{kopparapu2016inner,  way2016venus,turbet2016habitability,kumar2017habitable,boutle2017exploring,way2018climates,haqq2018demarcating,bin2018new,del2019habitable,sergeev2022trappist}, who found broadly consistent results, although the strength of the cloud feedback depends on details such as planetary rotation rate, stellar type, and the cloud pamaterization of the GCM used. 

Recently the concept of climate bistability near the inner edge of the HZ has been proposed \citep{turbet2021day, turbet2023water,selsis2023cool,chaverot2023first}. Their GCM simulations revealed that planets in this region might exhibit two stable climate states at the same forcing—a temperate climate and a hot, steamy climate—depending on initial conditions. They argue that a critical factor in this bistability is the role of nightside clouds, which can trap outgoing longwave radiation and contribute to warming, but they did not firmly establish this using techniques such as mechanism-denial experiments.

These studies have opened the possibility of exciting new climate behavior, but the mechanisms driving the bistability remain unclear, largely due to several unresolved challenges. First, a complete transition into the runaway greenhouse state is difficult to produce with state-of-the-art GCMs due to the inconsistency between the short timescales in the upper atmosphere and the longer timescales in the lower atmosphere \citep{turbet2021day}. Second, unconstrained assumptions in the sub-grid-scale parameterizations of GCMs can lead to widely varying results across models, especially when extrapolating to unknown conditions, as highlighted by \citep{yang2016differences,yang2019simulations}. Third, it is not always straightforward to uncover the underlying physics driving a mechanism in GCMs, which are inherently complex, computationally demanding, and difficult to analyze and fully understand.

The careful application of qualitative, low-order modeling can provide essential insight in a situation like this when the results from complex numerical models are relatively opaque \citep{abbot2011jormungand,abbot2016analytical,checlair2017no}. In particular, low-order models allow the flexibility to easily isolate and study specific mechanisms, providing clarity that can be obscured in GCMs \citep{held2005gap,schneider2006general,jeevanjee2017perspective}. In this study, we use a two-column, two-layer model to explore the interaction between the runaway greenhouse effect and cloud feedback mechanisms in driving climate bistability near the inner edge of the HZ. Our model, an updated version of that developed by \citet{abbot2009controls} and extended to exoplanets by \citet{yang2014low} (hereafter YA14), incorporates recent advances, including the effects of nightside clouds as described by \citet{turbet2021day} and \citet{chaverot2023first}. Additionally, our model includes an updated treatment of outgoing longwave radiation (OLR) under runaway greenhouse conditions. Using our refined model, we are able to establish that the bistability observed in GCMs is caused by the runaway greenhouse effect on the OLR in combination with either a dayside shortwave cloud feedback or a nightside longwave cloud feedback.

The paper is organized as follows: Section~\ref{sec:method} outlines our methodology, Section~\ref{sec:results} presents the results, and Section~\ref{sec:discussion} discusses the implications of our findings and their relevance to exoplanet habitability studies.

\section{Method} \label{sec:method}

\begin{figure}[ht!]
\epsscale{1}
\plotone{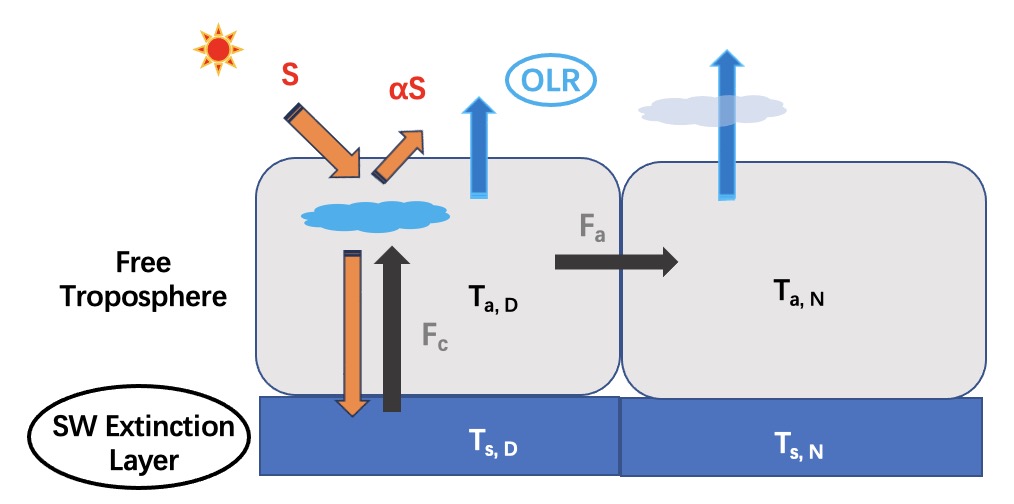}
\caption{Schematic representation of the two-column model. Gray boxes represent the free atmosphere (pressure \(< 1\) bar). Blue boxes represent the shortwave extinction layer (presure \(\sim 1\) bar). Orange arrows represent shortwave radiative fluxes, blue arrows represent outgoing longwave radiation to space, and black arrows represent atmospheric heat advection. The dayside column consists of a cloudy part and a clear-sky part; we also include an optional stratospheric (\(< 0.1\) bar) nightside cloud that blocks nightside longwave emission. The updated modules compared to \citet{yang2014low} are indicated by ovals.\label{fig:1}}
\end{figure}

Our low-order model builds on the two-column, two-layer approach described by \citet{yang2014low}, which itself is derived from the model of \citet{abbot2009controls}. This model aims to simulate the climate of a tidally locked planet by dividing the planet into two columns (dayside and nightside) and two atmospheric layers, resulting in four distinct boxes. Each box's energy balance is calculated to simulate the climate system. The YA14 model successfully replicates cloud and convective behavior near the inner edge of the HZ as produced by GCMs. For a detailed overview of the assumptions and equations in the YA14 model, see YA14. Here, we focus on revising the YA14 model to extend its applicability to a runaway greenhouse state. A schematic representation of our model, with revisions highlighted, is shown in Figure~\ref{fig:1}. 

The YA14 model defines the upper layer as the free troposphere and the lower layer as the boundary layer, where turbulent exchange with the surface is important. Under runaway greenhouse conditions, the atmosphere becomes more massive and optically thick, such that the absorbed shortwave flux decreases by two orders of magnitude from the top of the atmosphere to the 1-bar level, with the surface pressure significantly exceeding 1 bar \citep{turbet2023water}. Therefore, we redefine the lower layer as the shortwave extinction layer, the layer in which most of the downward shortwave radiation is absorbed, which may be far above the surface of the planet.

Regarding OLR, the YA14 model assumes a gray-gas approximation for the clear-sky portion of the free atmosphere, which cannot produce a runaway greenhouse effect in a two-layer model. To address this, we impose a transition to the runaway greenhouse state in the OLR of the model by fitting a curve to radiative transfer calculations in complex models \citep{yang2014low, yang2016differences, zhang2020does, turbet2021day, chaverot2022does, chaverot2023first, selsis2023cool}. As the temperature increases, the OLR initially follows the gray-gas approximation. It then transitions to emitting based on the constant emission temperature \(T_{\text{emis}}\), which marks the onset of the runaway greenhouse effect \citep{pierrehumbert2010principles,koll2018earth,jeevanjee2020cooling,jeevanjee2020simple}. After reaching \(T_{\text{emis}}\), the OLR plateaus until the temperature reaches \(T_{\text{end}}\), the point at which the runaway greenhouse process is complete. Beyond \(T_{\text{end}}\), the OLR begins to increase linearly with further temperature increases as follows:
\begin{equation}\label{eq:olr}
    OLR_{\text{clear}} =
    \begin{cases}
        \epsilon \sigma {T_a}^4  & T_a \leq  T_{\text{emis}} \\
        \epsilon \sigma T_{\text{emis}}^4 & T_{\text{emis}} < T_a \leq T_{\text{end}} \\
        C (T_a - T_{\text{end}})  & T_a > T_{\text{end}}
    \end{cases}
\end{equation}
where \(OLR_{\text{clear}}\) is the clear-sky OLR, \(\epsilon\) is the atmospheric emissivity, \(\sigma = 5.67\times 10^{-8}\, \mathrm{W/m^2/K^{4}} \) is the Stefan-Boltzmann constant, \(T_a\) is the free atmosphere's temperature, \(T_{\text{emis}} = 280\, \mathrm{K}\), \(T_{\text{end}} = 700\, \mathrm{K}\), and \(C = 3 \,\mathrm{W/m^2/K}\) is a fitting constant. The default \(OLR_{\text{clear}}\) presented in the main text does not include overshooting, which corresponds best to an atmosphere without a background non-condensate (e.g., Figure~1 in \citep{chaverot2022does}). We find that more sophisticated OLR functions, such as incorporating a transmission parameter for the pre-runaway OLR \citep{koll2018earth} or using a higher-order polynomial fit for the post-runaway OLR, do not qualitatively alter our conclusions about bistability. After the runaway greenhouse transition, the bulk atmosphere becomes much thicker (\(\gg 1\, \text{bar}\)) and stratified \citep{selsis2023cool}. The thickness and stratification level vary with volatile inventory and radiative transfer processes \citep{selsis2023cool}, making the correct values of \(C\) and \(T_{\text{end}}\) uncertain. The qualitative results we present here do not depend on such uncertainties. 

The YA14 model focuses on dayside clouds and does not include nightside clouds. However, \citet{turbet2021day, turbet2023water} and \citet{chaverot2023first} identified the presence of nightside clouds in their GCM simulations. They suggest that longwave heating from nightside, stratospheric clouds may be important for climate bistability. To test this idea, we impose a stratospheric longwave nightside cloud radiative forcing \(F_{\text{str}}\) based on GCM results from \citet{chaverot2023first}. The radiative forcing is represented as a function with two steps to capture the primary behavior observed in the data: it has no effect in temperate climates, provides constant heating in runaway greenhouse climates, and features a plateau with two linear transitions in between. The value of the plateau is not crucial, as no equilibrium solution exists within the temperature range it covers. Specifically,
\begin{equation}\label{eq:nc}
    F_{\text{str}} =
    \begin{cases}
        0 & T_{a} \leq 320 \, \mathrm{K} \\
        \frac{560}{11} (T_a - 320)F_{NC} & 320 \, \mathrm{K} < T_a  \leq 400 \, \mathrm{K} \\
        \frac{7}{11}F_{NC} & 400 \, \mathrm{K} < T_a \leq 700 \, \mathrm{K} \\
        \left[ \frac{7}{11} + \frac{400}{11} (T_a - 700) \right] F_{NC} & 700 \, \mathrm{K} < T \leq 800 \, \mathrm{K} \\
        F_{NC} & T > 800 \, \mathrm{K}
    \end{cases}
\end{equation}
where \(F_{NC} = 200 \, \mathrm{W/m^2}\) is the strength of the nightside cloud forcing. 

We also note the following slight differences between our model and that of YA14: 

1. The YA14 model assumes water vapor is dilute in the atmosphere (\(P_a \approx P_{c}\), where \(P_c\) is the mass of background gas in the free atmosphere). Here, we include the non-dilute limit: \(P_a = P_c + e\), where \(e\) is the vapor pressure based on the Clausius-Clapeyron relation. We set an upper limit for \(e\) of \(10^6\) Pa to avoid numerical issues. Overall, this revision makes the atmosphere approach the opaque limit slightly faster compared to the dilute assumption.

2. The \citet{abbot2009controls} model includes the possibility of both a stratified and convective atmosphere: It solves for the dayside convective energy flux \(F_c\) necessary to establish moist criticality (exact neutrality to moist convection) if the atmosphere is convectively unstable and sets \(F_c\) to zero if the atmosphere is convectively stable. YA14 only considered cases where the atmosphere was convective, but we will consider both stratified and convective atmospheres as did \citet{abbot2009controls}.

3. We assume no oceanic heat transport between the dayside and nightside following \citet{chaverot2023first}. 

With these assumptions and revisions, the low-order model is constructed by requiring energy balance in each of the four boxes (Equations~(\ref{eq:1})–(\ref{eq:4})), applying the weak temperature gradient (WTG) approximation (Equation~(\ref{eq:5})), and enforcing convective criticality on the dayside (Equation (\ref{eq:6})). The model equations are:
\begin{equation}\label{eq:1}
    (1-\alpha)S - F_c + (1-f_c)\epsilon_D \sigma T_{a,D}^4 + f_c \sigma T_{c}^4 - \sigma T_{s,D}^4 = 0   
\end{equation}
\begin{equation}\label{eq:2}
    F_{c} - F_a + (1-f_c)\epsilon_{D} \sigma T_{s,D}^4 + f_c\sigma T_{s,D}^4 - (1-f_c)(\epsilon_{D} \sigma T_{a,D}^4 + OLR_{\text{clear}}) - 2f_c\sigma T_{c}^4 = 0    
\end{equation}
\begin{equation}\label{eq:3}
    F_a - F_d + \epsilon_{N} \sigma T_{s,N}^4  -  \epsilon_{N} \sigma T_{a,N}^4 - OLR_{\text{clear}} + F_{\text{str}} = 0    
\end{equation}
\begin{equation}\label{eq:4}
    F_d + \epsilon_{N} \sigma T_{a,N}^4 - \sigma T_{s,N}^4 = 0   
\end{equation}
\begin{equation}\label{eq:5}
    T_{a,D} - T_{a,N} = 0  
\end{equation}
\begin{equation}\label{eq:6}
    MSE_{s,D} 
    \begin{cases}
        = MSE^*_{a,D} & \text{convective dayside, solve for \(F_{c}\)} \\
        < MSE^*_{a,D} & \text{stratified dayside, \(F_{c} = 0\)}
    \end{cases}
\end{equation}

In these equations, \(T_{s,D}\) and \(T_{s,N}\) are the dayside and nightside temperatures of the shortwave extinction layer, respectively; \(T_{a,D}\) and \(T_{a,N}\) are the dayside and nightside temperatures of the free troposphere, respectively; \(S\) is the stellar flux averaged over the dayside; \(\alpha\) is the planetary albedo; \(\epsilon_D\) and \(\epsilon_N\) are the emissivities of the dayside and nightside, respectively; \(T_c = 230\,\mathrm{K}\) is the dayside cloud emission temperature; \(f_c\) is the dayside cloud fraction; \(F_c\) is the convective heat flux from the dayside lower layer to the free atmosphere; \(F_a\) represents atmospheric heat transport from dayside to nightside; \(F_d = 0.2 F_a\) is the nightside heating that goes into the shortwave extinction layer due to adiabatic descent. Using Equations~(\ref{eq:1})–(\ref{eq:6}), we determine solutions for the six unknown variables of the model: temperature (\(T_{s,D}\), \(T_{s,N}\), \(T_{a,D}\), and \(T_{a,N}\)), horizontal heat transport (\(F_a\)), and dayside convective heat flux (\(F_c\)). This solution is a function of model parameters; here we focus on the strength of stellar flux (\(S\)).

\section{Results} \label{sec:results}

\begin{figure}[ht!]
\epsscale{1}
\plotone{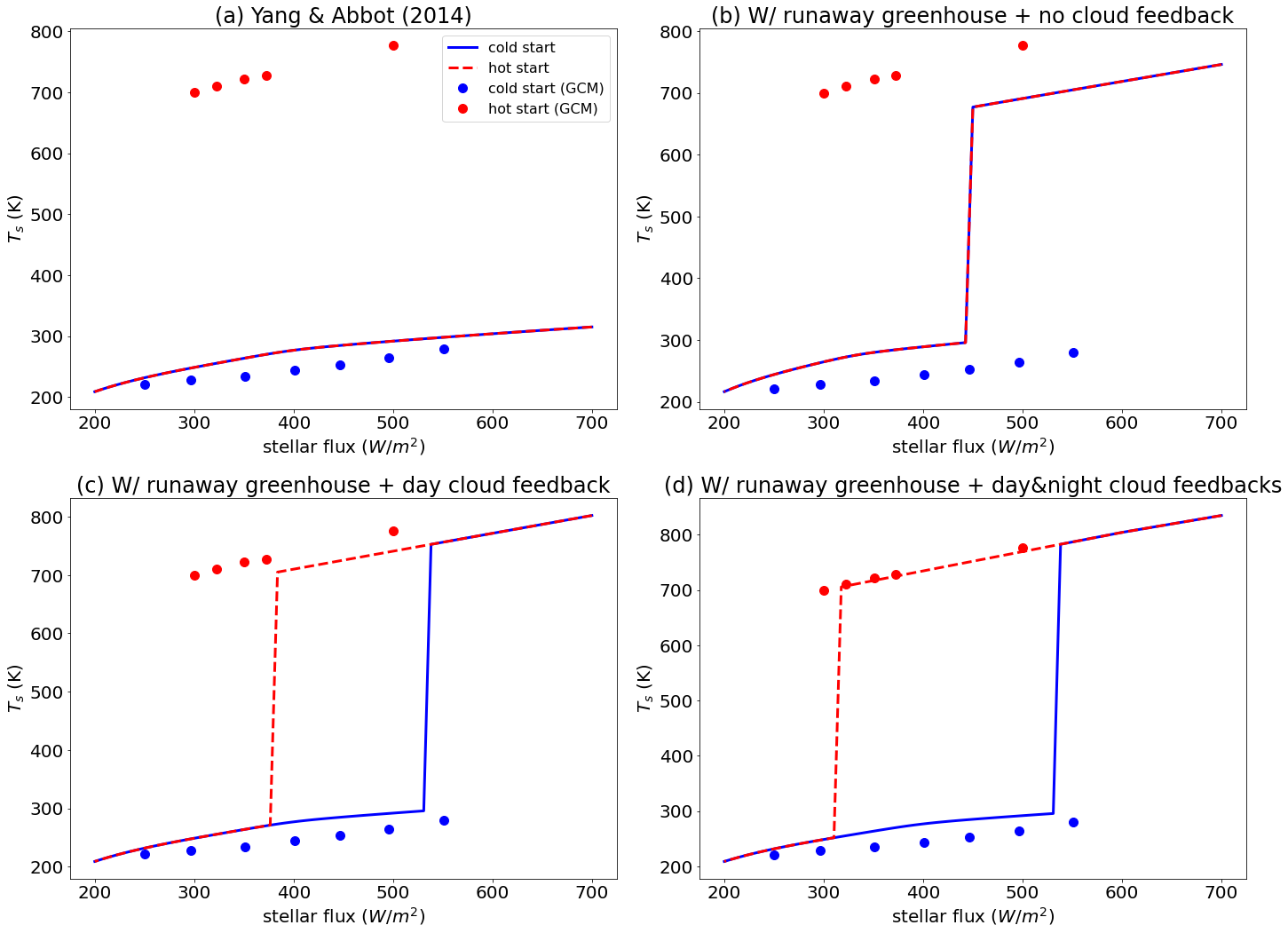}
\caption{Climate bistability at the inner edge of the habitable zone due to the runaway greenhouse effect. The figure shows the evolution of mean surface temperature with mean stellar flux. The blue results represent simulations with a cold start, and the red results represent simulations with a hot start. Solid and dashed lines represent results from our two-column model, while colored dots represent results from GCMs \citep{yang2013stabilizing,turbet2023water}. Panels show: (a) YA14 settings \citep{yang2014low}; (b) Addition of runaway greenhouse effect to YA14 with dayside cloud fraction fixed; (c) Addition of runaway greenhouse effect to YA14; (d) Addition of nightside cloud feedback to (c).\label{fig:2}} 
\end{figure}

We now investigate how various physical processes influence climate bistability using a step-by-step approach (mechanism denial experiments). Our baseline simulation, very similar to the YA14 model, does not exhibit bistability, consistent with the findings of YA14  (Figure~\ref{fig:2}a). The mean surface temperature changes smoothly with stellar flux and converges to a single solution, regardless of whether the simulation starts from a hot or cold state.

We next test a scenario in which we introduce the runaway greenhouse effect but fix the dayside cloud fraction to a constant, thereby disabling the dayside cloud feedback. This is similar to 1D radiative convective modeling in that the clouds are fixed \citep[e.g.,][]{kopparapu2013habitable, wordsworth2013water}, although our model includes separate dayside and nightside columns. In this scenario we find a jump in surface temperature as the stellar flux increases when the runaway greenhouse transition is encountered (Figure~\ref{fig:2}b), but we do not find climate bistability. 

When we include both the runaway greenhouse effect and the dayside cloud feedback, we observe climate bistability (Figure~\ref{fig:2}c). In the cold climate state the dayside atmosphere is convective, and therefore cloudy. As a result the albedo is high, allowing the cold state to exist. In the hot climate state the dayside atmosphere is stratified,  and therefore cloudless (according to our modeling assumptions). This leads to a low albedo and warming by absorption of stellar flux, allowing the hot state to exist. The bistability in our model is qualitatively similar to the behavior observed in GCMs, particularly for the cold branch. If we include a nightside cloud feedback as proposed by \citet{turbet2021day} and \citet{chaverot2023first}, but no dayside cloud feedback, we obtain climate bistability, although over an even smaller range of stellar fluxes (not shown). These results highlight the importance of both the runaway greenhouse effect and cloud feedbacks for producing climate bistability near the inner edge of the HZ. Disabling either mechanism prevents climate bistability in our model.

Finally, we show that including both dayside and nightside cloud feedbacks, in addition to the runaway greenhouse effect, improves the quantitative comparison between our model and the GCMs (Figure~\ref{fig:2}d). In particular, nightside cloud warming allows the hot state to exist at lower stellar fluxes, so that range of stellar fluxes for which climate bistability exists in our model is more similar to that in GCMs. 

\begin{figure}[ht!]
\epsscale{1}
\plotone{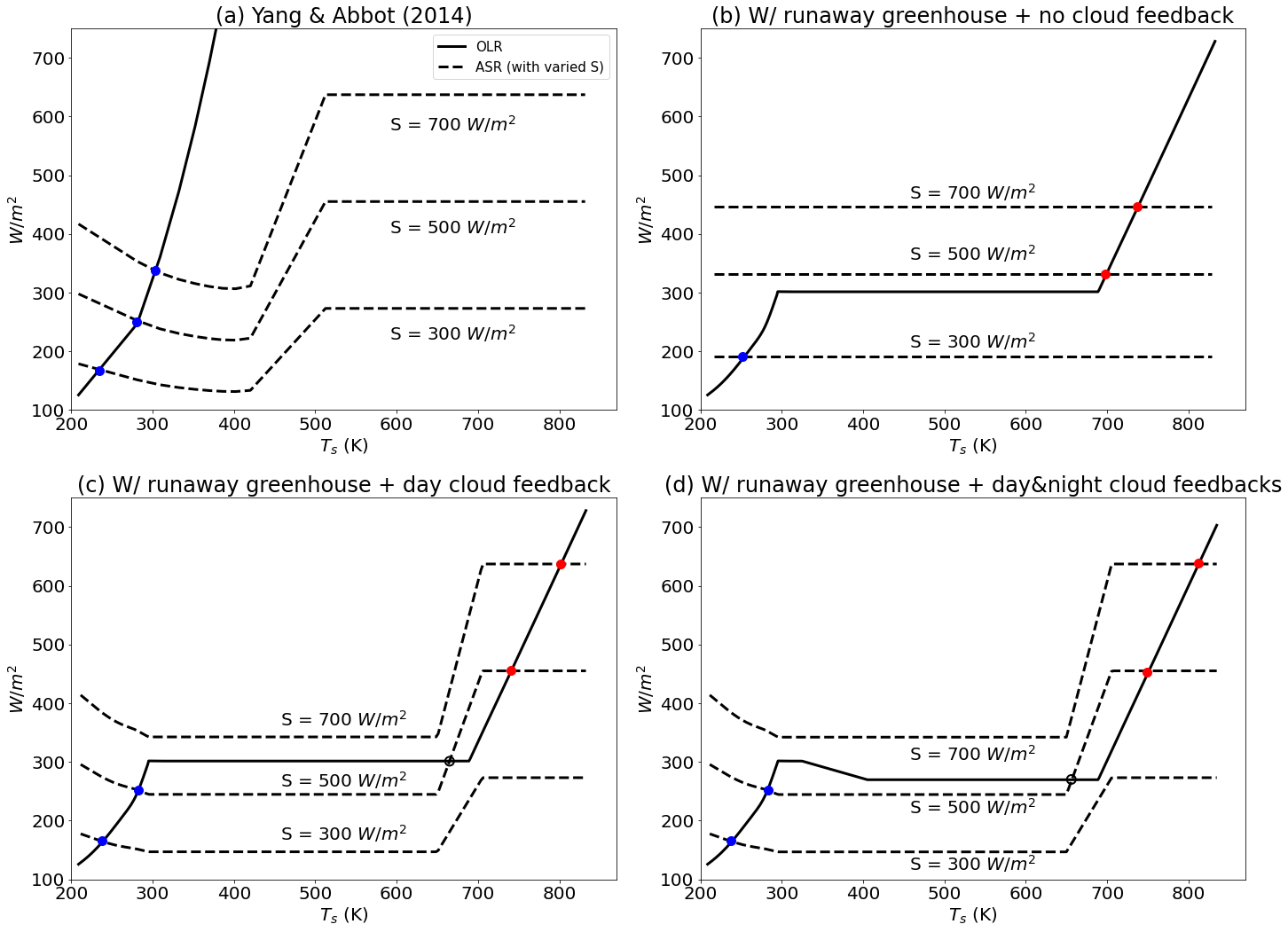}
\caption{The climate bistability can be understood by considering energy balance between outgoing longwave radiation (OLR, solid line) and absorbed stellar radiation (ASR, dashed lines). Each panel represents the corresponding simulation from Figure~\ref{fig:2}. Each intersection represents an equilibrium climate state, where blue closed circles represent stable cold climates, red closed circles represent stable hot climates, and the open circle represents an unstable climate during the runaway greenhouse transition.\label{fig:3}} 
\end{figure}

The climate bistability can be understood through the lens of energy balance. Figure~\ref{fig:3} illustrates how global mean outgoing longwave radiation (OLR) and absorbed stellar radiation (ASR) vary with global mean surface temperature (\(T_s\)) and stellar flux (\(S\)). Intersections of the OLR and ASR curves are fixed points, or climate solutions, that correspond to energy balance and may or may not be stable to small perturbations. Climate bistability requires two stable fixed points at the same external forcing (in this case stellar flux). Nonlinearity in the OLR and/or ASR curves is required to allow multiple intersections at the same stellar flux. The runaway greenhouse effect is the main source of this nonlinearity for the OLR curve, although nightside clouds also contribute. Specifically, the surface temperature increases by almost 400~K at the single OLR corresponding to the runaway greenhouse. The dayside cloud feedback supplies nonlinearity to the ASR curve. If the temperature is lower than about 650~K the dayside atmosphere is convecting, there is ample cloud cover, the albedo is high, and the ASR is low \citep{yang2013stabilizing}. If the temperature is higher than about 650~K atmospheric convection ceases on the dayside, there are no convective clouds, the albedo is low, and the ASR is high \citep{chaverot2023first}. The shapes of the OLR and ASR curves as various processes are included or excluded (Figure~\ref{fig:3}) provides a simple graphic explanation for why climate bistability near the inner edge of the HZ requires both the runaway greenhouse effect and one of the cloud feedbacks. Including the dayside cloud feedback, but not the runaway greenhouse effect, (Figure~\ref{fig:3}a) allows for the possibility of multiple intersections at very high stellar flux, but not near the inner edge of the HZ. Including the runaway greenhouse effect, but no cloud feedbacks, (Figure~\ref{fig:3}b) does not produce multiple intersections. It is only when we include both the runaway greenhouse effect and the dayside cloud feedback (Figure~\ref{fig:3}c) that the OLR and ASR curves acquire shapes that allow multiple intersections in the relevant stellar flux range. As can be seen in Figure~\ref{fig:3}d, the OLR has an overshoot when the nightside cloud feedback is included, which allows for a smaller range of climate bistability even without the dayside cloud feedback (constant ASR curves as in Figure~\ref{fig:3}b). Moreover, nightside clouds decrease the OLR in the runaway state (Figure~\ref{fig:3}d) allowing the improved fit to the GCMs in the hot state observed in (Figure~\ref{fig:2}d).

\begin{figure}[ht!]
\epsscale{1.1}
\plotone{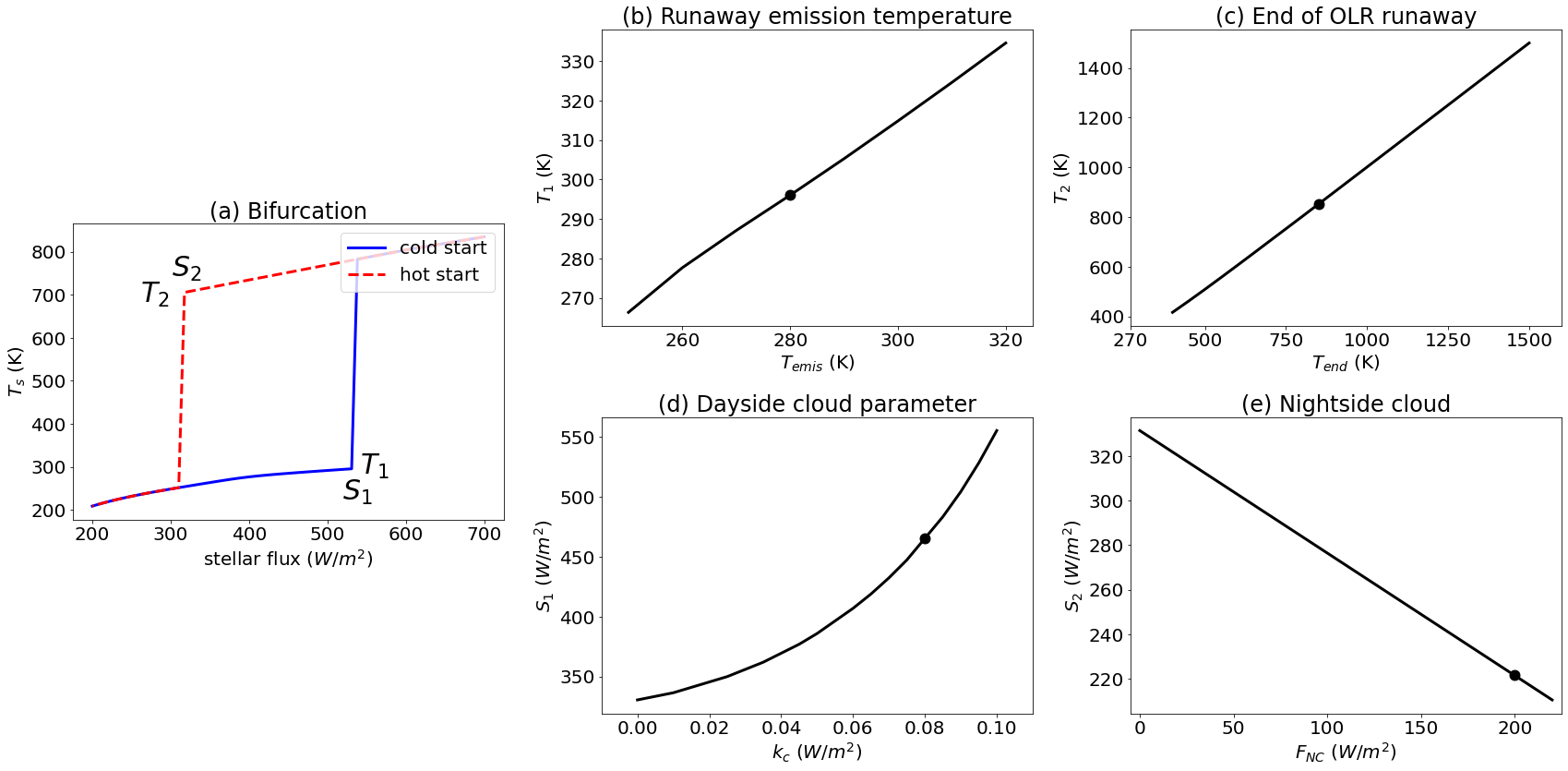}
\caption{The size of bistability is controlled by the runaway greenhouse effect and cloud parameters. For each parameter, the base case (panel a, same as Figure~\ref{fig:2}d) is adjusted. The change in bistability size is quantified by the temperature (\(T_1\)) and stellar flux (\(S_1\)) at which the model transitions from the cold state to the hot state (runaway greenhouse) as well as the temperature (\(T_2\)) and stellar flux (\(S_2\)) at which the model transitions from the hot state to the cold state (runaway condensation). \(T_1\) and \(T_2\) are controlled by \(T_{\text{emis}}\) (panel~b) and \(T_{\text{end}}\) (panel~c), respectively, the temperatures at which the runaway greenhouse process begins and ends. \(S_1\) and \(S_2\) are controlled by \(k_c\) (panel~d) and \(F_{NC}\) (panel~e), respectively. \(k_c\) describes the relationship between convection and cloud fraction while \(F_{NC}\) is the decrease in outgoing longwave radiation on the nightside if the nightside is completely covered with clouds. Circles indicate parameter values in the base case (Figure~\ref{fig:2}d).\label{fig:4}} 
\end{figure}

Since the climate bistability is driven by the runaway greenhouse effect and cloud feedbacks, model parameters associated with these processes exert primary control on bistability characteristics (Figure~\ref{fig:4}). We quantify characteristics of the bistability by the runaway greenhouse temperature (\(T_1\)), runaway greenhouse stellar flux (\(S_1\)), runaway condensation temperature (\(T_2\)), and runaway condensation stellar flux (\(S_2\)) (Figure~\ref{fig:4}a). We find that \(T_1\) and \(T_2\) are primarliy controlled by runaway greenhouse parameters, whereas \(S_1\) and \(S_2\) are primarily controlled by cloud feedback parameters. Increasing the runaway emission temperature (\(T_{\text{emis}}\)) increases the temperature range over which the cold state is able to exist (decreasing the range of temperature bistability), which corresponds to increasing \(T_1\) (Figure~\ref{fig:4}b). Increasing the end-of-runaway greenhouse temperature (\(T_{\text{end}}\)) increases the minimum temperature at which the hot, runaway state stabilizes (increasing the range of temperature bistability), which corresponds to increasing \(T_2\) (Figure~\ref{fig:4}c). Increasing the strength of the dayside cloud feedback (\(k_c\)) increases the albedo in the cold state which allows the cold state to exist at higher stellar fluxes (increasing the range of bistability in stellar flux), which corresponds to higher \(S_1\) (Figure~\ref{fig:4}d). Increasing the amount of OLR blocked by nightside clouds (\(F_{NC}\)) warms the hot state and allows it to exist at lower stellar fluxes (increasing the range of bistability in stellar flux), which corresponds to lower \(S_2\) (Figure~\ref{fig:4}d). If we continue to decrease the runaway greenhouse effect by reducing the length of the plateau, or diminish cloud feedback by lowering cloud parameters to zero, the bifurcation gradually disappears (not shown), consistent with our mechanism-denial experiments.

\section{Summary and Discussion} \label{sec:discussion}

Our study uses a two-column, two-layer model to explore and explain the climate bistability observed at the inner edge of the habitable zone (HZ) in GCMs. This bistability is characterized by two distinct climate states: a cold, dry atmosphere and a hot, steamy atmosphere. We can only reproduce the climate bistability by incorporating both the runaway greenhouse effect and cloud feedbacks. In the cold state there is atmospheric convection on the dayside, convective clouds, and as a result, a high albedo that sustains lower temperatures. In the hot state the atmosphere becomes stable to convection so there are no convective clouds. As a result the albedo is low, allowing the hot state to exist. Nightside stratospheric clouds in the hot state provide additional warming, allowing the hot state to exist at a lower stellar flux than it otherwise would.

By the nature of low-order, conceptual modeling, our work neglects many potentially important effects, such as cloud microphysics \citep{yang2024impact}, small-scale cloud behavior \citep{sergeev2020atmospheric,lefevre20213d,yang2023cloud,sergeev2024impact}, water vapor buoyancy \citep{yang2020incredible,seidel2020lightness,yang2022substantial}, and nonlinear limit cycles in precipitation as the runaway greenhouse threshold is approached \citep{seeley2021episodic,dagan2023convection,spaulding2023emergence,song2024critical,yang2024predatorpreyminimumrecipe}. Nevertheless we are able to explain the climate bistability identified in complex GCMs by \citet{turbet2023water} with a simple mechanism involving the runaway greenhouse effect and a cloud feedback on either the day or night side. It is unlikely that the inclusion of more complex physical effects would alter this basic qualitative insight.

An important assumption of our model is the separation between a dayside column and a nightside column, which is appropriate in the limit of a tidally locked planet \citep{yang2014low,pierrehumbert2019atmospheric}. This orbital configuration is relevant for many planets near the inner edge of the HZ since they are close to their star and therefore subject to strong tidal interactions. Given the correspondence between our results and those in GCMs simulating planets that are not tidally locked, we expect our main conclusions to be robust to other orbital configurations. Future work could extend our model to faster rotation by incorporating a meridional Ekman mass transport in the free atmosphere as in Equation~8 of \citet{ding2018global}. Moreover, recent modeling efforts have found that a strong day-night contrast is a robust feature in runaway greenhouse atmospheres, regardless of orbital configuration \citep{turbet2019runaway,chaverot2023first}. This is probably because for such hot atmospheres the radiative timescale is shorter than the dynamical timescale \citep{koll2016temperature,haqq2018demarcating}. This suggests that our current modeling assumptions may be appropriate in the hot state even for fast rotating planets.

Another important assumption in our work is the absence of dayside clouds when the dayside atmosphere is stratified, since we only consider convective clouds on the dayside. Stratus clouds, which are a significant feature on Earth \citep{yu2004climate,wood2015stratus}, could also be relevant in the hot, non-convecting climate state. If so, the bistability mechanism should still operate as long as the albedo remains higher with convective clouds in the cold state than with stratus clouds in the hot state. Even if this condition is not met, it is possible that nightside stratospheric clouds could still generate bistability. Future work should aim to better understand the role of stratus clouds in the hot state using simple models, GCMs, and convection-permitting models \citep{spaulding2023effects,chaverot2023first,yang2023cloud}.

Our study is consistent with the recent suggestion that rocky planets near the inner edge of the habitable zone may remain in a hot, puffy state, exhibiting observable runaway greenhouse-induced radius inflation effects \citep{turbet2019runaway,turbet2023water}. This phenomenon, marked by a significant increase in planetary radius due to an optically thick, water vapor-dominated atmosphere, could be detectable with current and upcoming space missions, such as JWST. Detecting this radius inflation would provide a crucial observational diagnostic for probing the presence of water on Earth-sized exoplanets and testing the boundaries of the HZ concept.

Our results suggest a novel regime of climate bistability that may apply more broadly than for terrestrial planets near the inner edge of the HZ. For example, the interplay between cloud radiative feedbacks and the runaway greenhouse effect has been discussed for thick exoplanetary atmospheres \citep{tan2019atmospheric, tan2021convection}, particularly for giant exoplanets and brown dwarfs. Additionally, runaway greenhouse processes can occur with other condensable species \citep{pierrehumbert2010principles,koll2018earth}. For example, lava planets like 55 Cnc e \citep{demory2016map, hammond2017linking, zilinskas2022observability, hu2024secondary, patel2024jwst} may harbor thick atmospheres with silicate vapor, potentially experiencing bistability if radiatively active clouds are present. With the advent of JWST, there is an unprecedented opportunity to observe the atmospheres of ultra-hot exoplanets.  Future work should expand our model to consider a broader range of stellar types and atmospheric compositions, enhancing our understanding of exoplanetary climates across different environments.

\begin{acknowledgments}
This work was supported by NASA award No. 80NSSC21K1718, which is part of the Habitable Worlds program. B.F thank Wencheng Shao, Ziwei Wang, Xuan Ji, Leslie Rogers, and Tad Komacek for helpful discussions.
\end{acknowledgments}

\bibliography{sample631}{}
\bibliographystyle{aasjournal}

\end{document}